# Spectral properties and geology of bright and dark material on dwarf planet Ceres

G. Thangjam[1], A. Nathues[1], T. Platz[1], M. Hoffmann[1], E. A. Cloutis[2], K. Mengel[3], M. R. M. Izawa[4], D. M. Applin[2]

[1]Max Planck Institute for Solar System Research, Justus-von-Liebig-Weg 3, 37077 Goettingen, Germany (thangjam@mps.mpg.de, nathues@mps.mpg.de, platz@mps.mpg.de, hoffmann@mps.mpg.de),

[2]University of Winnipeg, Winnipeg, MB R3B 2E9, Canada (e.cloutis@uwinnipeg.ca, daniel.m.applin@gmail.com),

[3]IELF, TU Clausthal, Adolph-Roemer-Straße 2A, 38678 Clausthal-Zellerfeld, Germany (gkmengel@t-online.de),

[4]Institute for Planetary Materials, Okayama University – Misasa, TU 827 Yamada, Misasa, Tottori 682-0193, Japan (matthew.izawa@gmail.com)

Corresponding author: thangjam@mps.mpg.de

**Abstract**

Variations and spatial distributions of bright and dark material on dwarf planet Ceres play a key role for the understanding of the processes that have led to its present surface composition. We define limits for "bright" and "dark" material in order to distinguish them consistently, based on the reflectance with respect to the average surface using Dawn Framing Camera data. A systematic classification of four types of bright material is presented based on their spectral properties, composition, spatial distribution, and association with specific geomorphological features. We found obvious correlations of reflectance with spectral shape (slopes) and age; however, this is not unique throughout the bright spots. Although impact features show generally more extreme reflectance variations, several areas can only be understood in terms of inhomogeneous distribution of composition as inferred from Dawn Visible and Infrared Spectrometer data. Additional materials with anomalous composition and spectral properties are rare. The identification of the origin of the dark, and particularly the darkest, material remains to be explored. The spectral properties and the morphology of the dark sites suggest an endogenic origin, but it is not clear whether they are more or less primitive surficial exposures or excavated sub-surface but localized material. The reflectance, spectral properties, inferred composition, and geologic context collectively suggest that the bright and dark materials tend to gradually change towards the average surface over time. This could be because of multiple processes, i.e., impact gardening/space weathering, and lateral mixing, including thermal and aqueous alteration, accompanied with changes in composition and physical properties such as grain size, surface temperature, porosity (compaction).

## 1.0 Introduction

Dwarf planet Ceres (diameter ~940 km) is the largest and most massive object in the main asteroid belt, located at a mean heliocentric distance of ~2.8 AU. Ceres is likely a partially evolved carbonaceous chondrite-like body that underwent differentiation to some degree (Park et al. 2016; McCord and Castillo-Rogez, 2017). The aqueously altered carbonaceous chondrites of the CI and CM classes are the closest known analogue material of Ceres, in particular the CI chondrite-like material that underwent substantial aqueous alteration though there is no compelling evidence for any known meteorite to have been delivered from Ceres (McSween et al. 2017; Prettyman et al. 2017). Small halite-dominant grains in the ordinary (H) chondrites Zag and Monahans have been speculated to have originated on Ceres, and notably contain inclusions of carbonaceous chondrite-like materials (e.g., Fries et al., 2013; Zolensky et al., 2015, 2016). The large quantities of ammoniated phyllosilicates inferred from telescopic data (King et al., 1992) and VIR reflectance spectra (De Sanctis et al., 2016) are unknown in any carbonaceous chondrites (or from any other meteorite), nor from the possible cerean silicates hosted in halite inclusions within H chondrite breccias (Fries et al., 2013; Zolensky et al., 2015, 2016).

The dark carbonaceous chondritic-like surface of Ceres was subjected to aqueous and thermal alteration processes and mixed with bright and compositionally diverse subsurface materials (e.g., Nathues et al. 2016, 2017a,c; McCord and Castillo-Rogez 2017; McSween et al. 2017). Recent investigations suggest evidence for surficial water-ice in a few localized regions on Ceres (Combe et al. 2016, 2017; Prettyman et al. 2017; Nathues et al. 2017b; Platz et al. 2017). In addition, potential ongoing sublimation activities have been suggested (Nathues et al. 2015a; Thangjam et al. 2016b) though it is not yet confirmed (Schroeder et al. 2017). Gravity data from Dawn suggest ~30% of bulk water/water-ice content of Ceres (Park et al. 2016; McCord and Castillo-Rogez 2017; McSween et al. 2017) and Gamma Ray and Neutron Detector (GRaND) observation led to the detection of extensive near-surface water-ice, particularly in the higher latitudes (Prettyman et al. 2017). Investigating the composition and the physical properties of cerean surface features is important to understand the geological activities that shaped the present surface. Thus, a survey of the prominent bright and dark materials could shed light into acting geologic processes, since these materials are expected to provide insights into the evolved/processed regolith from likely pristine chondritic material.

Nathues et al. (2015a) published the first geologic map displaying the locations of bright material on Ceres using data acquired by the Framing Camera (FC), the imager onboard NASA's Dawn spacecraft (Sierks et al., 2011). Stein et al. (2017) presented results on the formation of bright material (faculae) and their geologic context using FC clear filter data, and Palomba et al. (2017) focused on studying the compositional differences and the evolutionary path among the bright spots using infrared (IR) data from the Visible and Infrared Spectrometer (VIR). The analyses by Stein et al. (2017) and Palomba et al. (2017) use a relatively similar approach but different methods to identify bright spots, i.e., a bright spot is a site that differs by +30% in their reflectance or albedo with respect to the surrounding area. In our present study, we define bright and dark material by using absolute reflectance limits with respect to the average Ceres surface, and present their spectral properties and geologic context by using FC color and VIR data.

## 2.0 Data and methodology

The FC instrument is equipped with one clear filter and seven color filters in the wavelength range from 0.44 to 0.96 µm (Sierks et al. 2011). Color filter data were used for compositional (mineralogical) studies, while the clear filter data were primarily used for geologic mapping and digital terrain modelling. So far more than 57,000 images of Ceres have been obtained by FC including a global coverage (color and clear filter) at a resolution of ~140 m/pixel during HAMO (High Altitude Mapping Orbit) and XMO2 (Extended Mission Orbit 2) phase. A global coverage of high-resolution images of clear filter at ~35 m/pixel was also achieved during LAMO orbit (Low Altitude Mapping Orbit) including few selected sites in seven color filters.

The FC data product is processed in different levels: 1a, 1b and 1c, in which the final product is level 1b for clear filter and level 1c for color filters. Details of the processing pipeline of the data products are described in Nathues et al. (2014, 2015b). The radiance data (level 1b or 1c) are further converted to reflectance (I/F) taking into account the solar irradiance from a normally solar-illuminated Lambertian disk. Color cubes are processed using an inbuilt MPS (Max Planck Institute for Solar System Research) image processing pipeline (Reddy et al. 2012; Nathues et al. 2014, 2015b). The color cubes are then photometrically corrected to standard viewing geometry of 30° incidence, 0° emission, 30° phase angle using Hapke parameters (Hapke 2012 and references therein). For this correction, the HAMO shape model derived from FC clear filter stereo images (Preusker et al. 2016) is used.

Diagnostic absorption features, due to specific cerean surface mineralogy, are generally outside the FC spectral range. In addition, FC color data are influenced by physical properties of the regolith. Therefore, FC color data are less capable of revealing the surface mineralogy of Ceres. Thus, we use VIR data for the study of the surface mineralogy of selected sites.

The VIR instrument hosts two detectors, VIS and IR, covering the wavelength ranges from 0.3-1.1 µm and 1.0 to 5.1 µm, respectively (De Sanctis et al. 2011). The calibrated radiance data (level 1B) of VIR instrument, archived at Planetary Data System/National Aeronautics and Space Administration (PDS/NASA, https://pds-smallbodies.astro.umd.edu/data_sb/missions/dawn/index.shtml) are used here. These data are converted to reflectance (I/F) data, and then applied thermal correction (Clark et al. 2011; Tosi et al. 2014; Nathues et al. 2017b). Thermal emission of the surface was removed by applying a method described in Clark et al. (2011) but adapted for the VIR spectral range. A boxcar average is applied for smoothing to increase the signal-to-noise ratio (S/N) of the spectra (e.g., Nathues et al. 2017b). However, since a photometric correction is not applied to the VIR data, a quantitative (spectral) analysis is not performed. Instead, we used the VIR data for a qualitative investigation, i.e., characterization of an absorption features and their relative band strengths.

VIR and FC data are complementary, the main differences being the covered wavelength ranges as well as the spatial and spectral resolution. Framing Camera data covers the range 0.44-0.96 µm, while VIR data are obtained from 0.3-5.1 µm, from which the range 1.0 to 4.1 µm is used here. The spatial resolution of FC imagery is approximately 2.5 times higher than VIR imagery. Thus, the high spatial resolution in combination with the high S/N of the FC imagery is used to derive additional spectral information and better link spectroscopic properties with geomorphology (Thangjam et al. 2013, 2014, 2016a, b; Nathues et al. 2014, 2015a, b, 2016, 2017a, b, c).

[Fig 1]

**3.0 Ceres bright and dark material**

Ceres' bright and dark materials are often defined on a relative scale (Stein et al. 2017; Palomba et al., 2017). Stein et al. (2017) identified bright spots (termed 'faculae') being

>30% brighter than the surrounding areas by using the Bond albedo derived from FC clear filter data. Similarly Palomba et al. (2017) defined bright spots using the equigonal albedo at 1.2 µm ("R30") from VIR data with bright regions having an R30 value greater than 30% of the cube average. However, a relative scale of identification with respect to the surrounding area is inconsistent compared to the average surface of Ceres. For example, Palomba et al. (2017) included bright spots whose albedo is lower than the Ceres average albedo (i.e., those which are located in a relatively dark background), while areas showing albedo higher than the Ceres average (those located in a bright background) are not identified as bright spots. Similarly Stein et al. (2017) considered areas with lower bond albedo than the average surface (0.034) as bright spots (0.032-0.5), while many areas with higher bond albedo than the average surface are not identified. A comparison of the map of bright material presented in this work (Figs. 1-2) with those of Stein et al. (2017) and Palomba et al. (2017) revealed significant differences (see subsequent discussion).

In the present study, the bright and dark sites are defined by using reflectance limits, i.e., in an absolute scale with respect to the average surface. A meaningful definition of "bright" and "dark" material on a global scale can be achieved based on a statistically significant distinction from the global average surface reflectance regardless of their spectral properties and geologic appearance. It is worth mentioning that literally bright or dark material identified during the early phase of Ceres orbits were done initially based on their reflectance, and the information on their composition, morphology and geologic formation were retrieved later on. In this study, a globally consistent approach for the determination of the thresholds of bright and dark material was preformed based on the histogram of the global HAMO mosaic at 0.55 µm. Extreme outliers in reflectance values, for example Occator crater's brightest pixels, and data from higher latitudes are excluded while computing this histogram (Fig. 1A). The histogram is split into two wings at the reflectance of the bin with the maximum number of pixels. Successively increasing the reflectance intervals of the two wings of the histogram from that maximum (Fig. 1A), we have calculated the standard deviation within those intervals separately for both wings (Fig. 1B). While the smallest of those intervals are dominated by the scatter of the average background, the transition to dark or bright material is characterized by an increasing trend of the sequence of the standard deviations (Fig. 1B). The standard deviation value at the transition of the average background value to dark or bright material of the distribution (Fig. 1B) is used to define the window with respect to the average value for the determination of the thresholds of bright or dark material.

Based on this approach using HAMO cycle 1 and cycle 5 FC color data at 0.55 µm, the lower limit of bright material is found to be 0.035, and the upper limit of dark material is 0.028, while the range for average surface is 0.028 to 0.035. Figure 1C shows a global map of the distribution of bright (in white), dark (in black), and average surface (in gray). The data from XMO2 orbit, i.e., particularly in the polar areas, are prone to topographic shadows and photometric inaccuracy enhancing the noise level, and therefore, analysis and interpretation of data from the high latitudes and polar areas need caution. The brightest material on Ceres is found on the Occator floor (reflectance >0.3) followed by sites at Oxo crater (reflectance ~0.15). The majority of the bright sites on Ceres show reflectance <0.1, and are associated with young impact craters (on crater rims, floors, and walls) and their ejecta. A closer inspection of LAMO data (~35 m/pixel) revealed that craters of a few hundred meters in diameter or less are frequently found superimposed on larger craters and it seems that these small impacts have re-excavated pre-deposited bright material (e.g., numerous small craters superimposed on Azacca crater and its ejecta). We mapped >50 craters with diameters >5 km and >1500 craters with <5 km diameter that exposed bright material. Figure 2 shows different views of the global mosaic of Ceres in Mollweide projection. The reflectance variation at 0.55 µm across the surface is displayed in Fig. 2A. An RGB color composite using R= 0.96 µm, G= 0.65 µm, B= 0.44 µm (Fig. 2B) illustrates the major variations in three FC colors. Another RGB composite of R=0.96/0.44 µm, G=0.65 µm, B=0.44/0.96 µm (Fig. 2C) shows the overall spectral slope in the FC wavelength range combined with the reflectance at 0.65 µm. The spectral shape of bright material is typically negative sloped longward of 0.55 µm (bluish color). Figure 2D displays the global RGB as in Fig. 2C but now superimposed with the mapped bright and dark sites. Green squares and yellow circles represent large and small craters showing bright material, respectively. Red stars indicate regions of dark material. This map shows that the majority of the bright spots are associated with the bluish areas. Bright material is mainly found in the equatorial and mid-latitudes, but some exist also in higher latitudes. Dark material is rare compared to bright material. Ejecta in the northeast of the Occator crater are the most prominent extent of dark material on Ceres. Other locations with notable extent of dark material are in the Nawish/Heneb crater region, Urvara crater floor and Yalode crater floor, while few patches of dark material are found in the southwestern Dantu ejecta and in some other sites (e.g., Consus crater, Gaue crater).

[Fig 2]

### 3.1 Bright material
Before going into a detailed analysis of bright material, it is important to mention that the catalogues of bright spots compiled by Stein et al. (2017) and Palomba et al. (2017) differ from ours due to the different approaches used. Stein et al. (2017) classified the bright spot/faculae into four types based on their geologic/morphologic setting: (1) those located at large crater central pits or peaks or crater floor fractures, (2) those located on crater rims and walls, (3) those located on ejecta, and rims of small craters in ejecta blankets, and (4) those associated with the unique surface feature Ahuna Mons. Stein et al. (2017) identified more than 300 faculae on global scale, where the majority of faculae (>200) are located on crater rims and walls. Palomba et al. (2017) listed 92 bright spots (BS) and classified them into four families based on their morphology: (1) ejecta; (2) crater associated features (CR- Crater Rims, CF- Crater Floors, CC- Crater Chain, CP- Crater Peak, or combination of them); (3) linear features. They also divided the bright spots (BS) into four groups based on their composition: (1) typical BS (similar composition with the average surface, i.e., low reflectance, Ca-Mg type of carbonate and low abundance), (2) Occator-like BS (brightest BSs, greatest abundance of carbonates, and sodium-rich carbonate and more OH-rich, Al-bearing phyllosilicate rather than Mg-bearing), (3) Oxo-like BS (moderate to high albedo, depleted in OH- and $NH_4$-bearing compounds, carbonate composition from Mg/Ca and/or Na, (4) Haulani-like BS (intermediate property between Occator-like BS and Oxo-like BS). A comparison of our map of bright material (Figs. 1, 2) with those of Stein et al. (2017) and Palomba et al. (2017) reveal significant differences. We identified more than 50 craters (diameter >5 km) with notable spatial extent of bright material (locations and feature names provided as SOM/catalogue), and >1500 craters which excavated bright material (diameter <5 km, majority of them in the order of hundreds of meters diameter). In addition, we found isolated bright spots and numerous small bright spots superimposed on bright ejecta of larger craters. The apparent inconsistencies among the three maps can be explained by the different approaches used for defining the bright material and their units: (1) definition of the thresholds, i.e., a relative scale with respect to the surrounding as adopted in Stein et al. (2017) and Palomba et al. (2017) versus or an absolute scale as adopted in this work; (2) identification and extent of the unit (for example, Azacca crater and its ejecta as one unit, or numerous smaller impact craters with bright material as different units which are superimposed on the crater or its ejecta).

Most of the bright material occurrences identified in this study are associated with impact craters, located on crater floors, rim/walls, and ejecta, but a few exceptions exist where bright material is associated with mons or volcanic dome-like features (e.g., Ahuna Mons; Ruesch et al. 2016) or impact cratering on such features. Some bright material shows unique spectral and compositional properties (e.g., Occator crater, Ernutet crater; De Sanctis et al. 2016, 2017; Nathues et al. 2016, 2017a, c) despite their association with impact craters. Therefore, a systematic study of bright material requires consideration of the geologic context and formation combined with spectral information. Figure 3 shows a flow chart of the varieties of bright material on Ceres. This approach divides bright material initially into two major groups depending on their association with impact or non-impact processes. Bright material clearly associated with impact processes are further divided into two classes depending on their formation with respect to the cratering event, i.e., whether the bright material was formed during the impact (syn-impact) or after the impact (post-impact). The syn-impact bright material is subdivided into three sub-groups based on their age, i.e., their morphologic appearance: fresh (e.g., Oxo, Haulani, Juling, Kupalo etc.), moderately degraded (e.g., Dantu, Azacca, Ikapati, etc.), and degraded (e.g., Ezinu, Messor, Kerwan, Urvara, etc.). The post-impact bright material is subdivided into two classes depending on their source or origin, i.e., whether the origin is endogenic (e.g., Occator bright material), or potentially exogenic (e.g., Ernutet bright material, i.e., material delivered by impactors (Nathues et al., 2016, 2017a; De Sanctis et al. 2016, 2017; Pieters et al, 2017).

[Fig 3]

Occator's bright material at Cerealia Facula and Vinalia Faculae is unique due to **its** extreme reflectance and composition (Nathues et al. 2017a, c; De Sanctis et al. 2016; Palomba et al., 2017; 'Occator-type' hereafter). Their origin by post-impact endogenic processes (Nathues et al. 2017a, c) of ascending subsurface brine is likely. Cerealia Facula is a central pit hosting a dome in its center while Vinalia Faculae is a cluster of bright spots to the north-east of Cerealia Facula (Nathues et al. 2015, 2017a, b, c). The bright material at the dome of Cerealia Facula shows overall red sloped spectra in FC wavelengths (Nathues et al., 2017a, c). Patches of less bright material than at the Faculae are also common at other places in Occator, i.e., on the crater floor, on crater rim/wall and ejecta.

Further unique examples of bright material are associated with the crater Ernutet (Nathues et al., 2016; De Sanctis et al. 2017; 'Ernutet-type' hereafter). Within and close to Ernutet, several sites exhibit entirely red-sloped FC color spectra due to the presence of organic-rich material (Nathues et al. 2016; Schroeder et al. 2017; De Sanctis et al. 2017; Pieters et al. 2017). Most of the organic-rich sites are bright material sites; however, not all bright material in this region shows entirely red sloped spectra in FC data or organics in VIR data.

All bright and dark material described so far is related to impact cratering processes, but Ahuna Mons is a volcanic dome formed by extrusion and inflation of cryomagma (Ruesch et al. 2016; Platz et al. 2017; 'Ahuna-type' hereafter). There is bright material associated with other mons features and impacts on these, for example, small to medium size craters (~5-20 km in diameter) that exposed bright materials at Liberalia Mons.

Figure 4 displays several RGB (R = 0.96 µm, G = 0.65 µm, B = 0.44 µm) mosaics of selected sites, i.e., fresh (A-C), moderately degraded (D-F), degraded (G-I) craters, as well as Occator (J), Ernutet (K), and Ahuna Mons (L). The RGB mosaics are stretched at a same scale, and thus color variations are comparable (a figure with different stretches is provided as SOM/S1).

[Fig 4]

The youngest craters (Fig. 4 A-C) exhibit the most prominent bright material on Ceres in terms of spatial extent and reflectance (e.g., Haulani, Oxo, Kupalo, Juling). Crater counting revealed ages of only few Ma for these craters: Oxo ~0.19 Ma/LDM (Nathues et al. 2017b), ~3 Ma/LDM, ~4 Ma/ADM (Hughson et al. 2017), ~0.5 Ma ADM/LDM (Schmedemann et al. 2016); Haulani ~1.7-2.7 Ma LDM, ~2.8-5.9 Ma ADM (Schmedemann et al. 2016). The ADM and LDM ages represent asteroid derived and lunar derived models, respectively (Hiesinger et al. 2016; Schmedemann et al. 2016). Typical of these craters are fresh ejecta and bright deposits on crater rims, walls and floors. Figure 5A and 5E show color spectra of these sites. The FC color spectra of bright sites at these small to medium sized craters are characterized by a reflectance maximum at 0.55 µm and a blue (i.e., negative) slope longward. However, we found some slight deviations: some of the bright sites exhibit their peak reflectance at 0.44 µm and are entirely blue-sloped (e.g., Oxo spectrum in Fig. 5A/E). In addition, a few locations show slightly positive slopes longward of 0.65 µm, as for example, portions of the southern floor of Haulani and Juling (not shown). In order to allow a comparison with the global average, each subfigure in Fig. 5 displays the global average cerean spectrum as well.

Bright sites appearing in moderately degraded craters are intermediate in terms of their reflectance between the fresh and heavily degraded craters. Their bright ejecta fields are more eroded compared to the fresh craters. However, superimposed small craters of primary or secondary origin that excavated the pre-deposited bright material are more frequent than for the fresh craters. Dantu is one of the large medium age impacts (~126 km in diameter) while Azacca and Ikapati craters are of medium size (~50 km in diameter). The crater size frequency distribution (CSFD) analysis of the ejecta and the crater floor gives an age range of ~25-28 Ma/ADM and ~72-150 Ma/LDM for Dantu (Williams et al. 2017; Kniessl et al. 2016), ~45 Ma ADM and ~75 Ma LDM for Azacca (Schmedemann et al. 2016), ~20-43 Ma /ADM and ~36-43 Ma/LDM (Pasckert et al. 2017), ~19-34 Ma/ADM and ~19-66 Ma/LDM (Schmedemann et al. 2016) for Ikapati crater. Figure 5B and 5F show color spectra of these craters. The FC color spectra are, as with the fresh craters, characterized by a reflectance maximum at 0.55 µm and a blue spectral slope longward. Also as in the case of the fresh craters, some of the bright sites exhibit their peak reflectance at 0.45 µm and are entirely blue sloped. Only a very few bright material sites at these medium age craters exhibit slightly positive spectral slopes longward of 0.65 µm (not shown here).

The heavily degraded craters comprise old craters showing eroded crater morphology and ejecta fields. Bright material exposures of mainly re-excavated material are present due to small fresh impact craters or distal ejecta from nearby large and young impacts. For example, many of the smaller craters with bright material at Ezinu (Fig. 3G), Messor (Fig. 3I), and Kerwan (Figs. 2C/D) seem to be associated with distal ejecta (secondary impact craters) from Occator and Dantu. Ezinu is one of the oldest large craters (diameter ~116 km) and the CSFD analysis gives an age range of 257-4120 Ma/ADM and 968-3380 Ma/LDM (Scully et al., 2017). Messor crater is a medium sized crater (~40 km), and its model age is 96 Ma/ADM and 192 Ma/LDM based on its ejecta (Scully et al. (2017), however, its morphologic appearance suggest an older age. The Kerwan basin (~280 km diameter) is one of the eroded basins on Ceres in which the topographic relief was smoothed due to viscous relaxation (Bland et al. 2016). The model age of Kerwan basin is ~280 Ma/ADM and ~1.3 Ga/LDM (Williams et al. 2017). Urvara basin (~140 km diameter) is smaller than Kerwan, and its surface topographic relief is not as relaxed as Kerwan's. The model age of Urvara basin is ~190-550 Ma/LDM and 73-120 Ma/ADM (Crown et al. 2017), ~110 Ma/ADM and ~242 Ma/LDM (Schmedemann et al. 2016). The FC color spectra of Urvara, Kerwan, and Ezinu are shown in Fig. 5C and 5G. The bright material spectra are similar to other bright areas on Ceres, i.e., peak reflectance at 0.55 µm and spectrally blue slope longward of the peak reflectance. A few locations show, within the measurement uncertainties, a slight red or neutral slope longward of the peak reflectance at 0.65 µm (not shown here).

The unusual sites (Fig. 4 J-L) exhibit unique spectral properties as for example, Occator, Ernutet, and Ahuna Mons. The crater counting suggest a wide range of ages (Occator ~4-34 Ma/LDM (Nathues et al. 2016, 2017a), ~19-21 Ma/LDM (Neesemann et al. 2017), Ernutet ~420 Ma/ADM and ~1.6 Ga/LDM (Pasckert et al. 2017), Ahuna Mons ~70 Ma/ADM and ~160-210 Ma/LDM (Ruesch et al. 2016). However, it is certain that the processes that exposed these bright materials are much younger than the surface features themselves because of their fresh morphologic appearance. Figures 5D and 5H show reflectance color spectra and normalized spectra at 0.65 µm of these unusual sites. Ahuna Mons bright material in FC wavelengths is spectrally similar to that of fresh crater' bright material, i.e., it is characterized by a reflectance maximum at 0.55 µm and blue-sloped longward of 0.55 µm. Spectral shapes similar to that of Ahuna Mons and other fresh craters are also found among bright material in Occator and Ernutet.

[Fig 5]

### 3.2 Dark material

We identified dark sites in multiple locations, and a few with significant spatial extent are displayed in Fig. 6 (locations and feature names provided as SOM/catalogue). To date the dark material is least explored in terms of composition and origin; its study could deliver insights into the primordial surface composition. It is speculated that carbon (usually found in association with organic matter and fine phyllosilicate matrices of CI/CM chondrites), magnetite and sulfides, which formed during aqueous alteration, could act as darkening agents (McSween et al. 2017; Pearson et al. 2006; Cloutis et al. 2011). However, assuming ~80% of darkening component while modelling VIR spectra seems to be an overestimation since GRaND observations suggest a much lower magnetite content of ~6% (McSween et al. 2017; Prettyman et al. 2017; De Sanctis et al. 2015). Interestingly, notable extents of dark material are observed in a few localized areas only (Fig. 6 A-D), though numerous areas of a few pixels in size of ambiguous identification have been found during the current study. Occator crater ejecta in the north/northeast/southeast show the largest spatial extent of dark material on Ceres (Fig. 6A). Smaller sites are in the Nawish/Heneb crater region (Fig. 6B), at Dantu crater ejecta (Fig. 6C) in the southwest, at Urvara (Fig. 6D) and Yalode (figure not shown) basins. The dark material on the floor of Urvara and Yalode is brighter than Occator dark material, and their spatial extent is less compared to that of Occator. Detailed mapping and investigation of dark material in terms of composition and morphology will be crucial to understand the geologic formation processes and the nature of the darkening effect acting on the cerean surface. Dark material at craters Nawish/Heneb seems to be associated with a pre-existing surface material that is likely different from other areas (e.g., Occator, Dantu) which are associated with their impact ejecta. It is obvious that dark material is found at different geologic settings and ages, for example, at the older Urvara and Yalode basins to the relatively younger Occator crater. Figure 6E and 6F display the reflectance spectra and the normalized reflectance spectra of the dark material compared with the global average spectrum of Ceres. All dark material spectra share a common spectral characteristic with the average surface, i.e., a peak reflectance at 0.65 µm. We found variations of the spectral slopes on either side of the peak reflectance, but the degree of these variations is negligible compared to the variations of the bright material. It is difficult to identify whether the dark material at Nawish/Heneb is either surficial dark material of primitive chondritic origin or impact excavated and then evolved/altered by impact-driven processes. However, dark material in the form of ejecta, for example, in Occator crater, implies either a localized subsurface dark material origin or an impact-related mechanism where the excavated material changes its compositional and/or physical properties and/or a shock-blackened mechanism was acting to resemble dark material. In addition, the Dantu crater shows bright and dark ejecta though not as prominent as for Occator.

[Fig 6]

## 4.0 Discussion

The overall analysis of Ceres bright and dark material from FC color data reveal common spectral properties among bright and among dark material. Bright material spectra (except the unusual ones as described above) show their peak reflectance at 0.55 µm and a blue spectral slope longward of this wavelength, while the peak reflectance for the dark material is at 0.65 µm and the spectra are blue sloped longward of 0.65 µm.

When compared to average Ceres, the bright component includes material that is brighter and more blue-sloped. The dark component, again compared to average Ceres, has a similar spectral shape but lower overall reflectance. The shift in wavelength of peak reflectance of bright (0.55 µm) to dark (0.65 µm) material is possibly due to the enrichment of the darkening component in the regolith, or lateral mixing with the darker background material. Bright and dark material spectra shown in Fig. 5 and 6 (except those of the unusual sites) are averaged to obtain "typical" Ceres bright and dark region spectra. Figure 7 displays these typical cerean bright and dark material spectra and compare them with the global average spectrum. The average bright material spectrum (typical-bright material, hereafter) exhibits a reflectance of ~0.07 at 0.55 µm, while this value is ~0.027 for the dark material. Figure 7B demonstrates that, as mentioned, the spectral shape of the average dark material spectrum is very similar to the global average, except a slight difference in the spectral slope shortward of the peak reflectance. Again, the bright material in general share similar spectral shape (with peak reflectance at 0.55 µm and blue slope longward), however, the variations in spectral slope on either side of the peak reflectance is stronger (Fig. 5E-G). Figure 7C and 7D display the typical bright material as well as the unusual bright material reflectance spectra of Occator- and Ernutet-type normalized at 0.96 µm.

[Fig 7]

In order to shed more light into the nature of bright and dark material, we compare in-flight spectral data with data of laboratory analogue samples in the next section.

### 4.1 Comparison with laboratory spectra

Because of its high spatial resolution and high S/N, the FC data is well-suited to reveal physical and compositional surface properties that affect the spectral shape and overall reflectance. Figure 8 compares spectral slopes and reflectance values of bright and dark sites with those of laboratory data of carbonaceous chondrites (CI, CM) and phyllosilicates (antigorite, one of the polytypes of Mg-dominant serpentine). The laboratory spectra have been resampled to FC bandpasses considering the response functions of the FC (Nathues et al., 2015a, b; Thangjam et al., 2013, 2014, 2016a, b). Lab spectra were compiled from the databases of RELAB (Brown University), and the Planetary Spectrophotometer Facility (University of Winnipeg).

The samples used here are Ivuna (CI1 chondrite), Orgueil (CI1 chondrite), Murchison (CM2 chondrite), and antigorite. Physical parameters and alteration processes that might influence the spectral properties of the surface include: (1) grain size, (2) finer-coarser mixture fraction/dominance, (3) heating temperature and aqueous alteration, and (4) porosity or compaction. The comparison is visualized by using two-dimensional plots of the overall slope (0.96/0.44 µm) versus reflectance at 0.55 µm (Fig. 8A-D) and the slope on either side of the 0.65 µm filter, i.e., 0.44/0.65 µm versus 0.96/0.65 µm (Fig. 8E-H). The 0.65 µm filter was chosen since it resembles the peak reflectance of the global average Ceres spectrum (e.g., Thangjam et al. 2016a).

The linear regression of bright material data points shown in Figure 8 excludes data points belonging to the unusual bright sites: those data points are plotted separately. For Fig. A-D, data points of Occator- and Ernutet-type lie outside the plotted reflectance and 0.96/0.44 µm slopes, and are not shown for clarity of the plots. Figure 8A and 8E highlight grain size influences on spectral shape by using lab samples (Orgueil: <45, <100, 40-100 µm; Murchison-1: <45, <90, 100-200, <200 µm; Murchison-2: <45, <90, <150, <250, <500, <1000 µm). Inflight data of the bright material do not follow the lab data trends as Figure 8A demonstrates, however, when plotting the parameter space 0.44/0.65 µm versus 0.96/0.65 µm a similar trend is found for in-flight and lab-data (Fig. 8E). In addition to individual grain size ranges, it is worth investigating potential influences of variable mixtures of finer and coarser grain-size fractions in the sample/regolith. Figures 8B and 8F show their influences on lab samples: (1) Murchison in 50% to 90% coarser fractions in 10% intervals (Izawa et al. 2016), (2) Antigorite sample in different fine and coarser fractions (grounded for several periods of time intervals of 1 min, 3 min, 10 min, 30 min; Bishop et al. 2008). Data of antigorite are shown in the inset. The inflight data and lab data in the parameter space of 0.96/0.44 µm versus reflectance at 0.55 µm do not show any obvious correlation (Fig. 8B), however the parameter space in 0.44/0.65 µm versus 0.96/0.65 shows a similar trend for the bright material (Fig. 8F). Fig. 8B and 8F exhibit similar dependencies as observed in Fig. 8A and 8E. Figures 8A-B and 8E-F demonstrate that the grain size influence on both spectral slopes (0.44/0.65 µm and 0.96/0.65 µm) of the in-flight data seems to be significant while its influence on reflectance is minor (lab samples show typically an increasing reflectivity with decreasing grain size). The latter could be due to a stronger albedo variation caused by composition masking grains size effects.

[Fig 8]

Figures 8C and 8G show sample heating influences on lab data. The Ivuna sample was heated from 100 to 700 °C in 100°C intervals (Hiroi et al. 1996), antigorite was heated from 400 to 1000°C in 100°C intervals but for two different grain sizes (0-65 and 65-125 µm, Hiroi et al. 1999), and Murchison was heated from 200 to 600 °C in 100 °C intervals (Izawa et al., 2016). The general trend measured on the laboratory samples in the parameter space 0.96/0.44 µm versus reflectance at 0.55 µm shows correlation with the in-flight data for the bright sites (Fig. 8C). The trends of lab data in 0.44/0.65 µm versus 0.96/0.65 µm parameter space are non-uniform, but overall the antigorite samples show a similar trend as the in-flight data for the bright material in this parameter space. Murchison-2 exhibits a slightly opposite trend compared to that of Murchison-1 and Ivuna sample. Nevertheless, from Figs. 8C and 8G, it is obvious that the reflectance of the lab samples increases with increasing heating temperature while the overall slope decreases. Thus differential heating of cerean regolith could also be responsible for the observed variations in spectral shape.

Figures 8D and 8H show the effect of the compactness (porosity) of the lab samples. Spectra of the Orgueil sample is from a "normal" and "tight" packing (from RELAB), while it is a regular-, loose- and tight packing for the Murchison sample (Izawa et al. 2016). The sample with the tight packing shows higher reflectance compared to the loose or regular packing samples, and the overall spectral slope also slightly lowered (Fig. 8D, H). However, it is difficult to compare with the inflight data because of paucity of lab sample/data; the lab data suggest that porosity or compactness might affect the reflectance more strongly than the spectral slope.

In general, for the carbonaceous chondrites, increasing grain sizes result in flatter spectral shape and lower reflectances (Figures 8A-B, E-F). Heating of carbonaceous chondrites results in flatter spectral shape and increasing reflectance (Figure 8C, G), while compaction results in a minor flattening of spectral shape and increasing reflectance (Fig. 8D, H). A detailed quantitative analysis and interpretation of these effects on spectral properties is subject of upcoming investigations. Having said this, it needs to be re-called that CI and CM chondrites are not actual cerean spectral analogues. Thus, caution is needed while interpreting and comparing the results of laboratory investigations of carbonaceous chondrites with inflight data.

Therefore, a gradual change of the spectral properties of the surface could be due to a change of physical properties and composition. The compositional variation of bright and dark material is presented in the following section.

**4.2 Bright and dark material composition**

A detailed analysis of compositional differences among bright spots on Ceres is presented by Palomba et al. (2017), and a global analyses by others (e.g., De Sanctis et al. 2015; Ammannito et al. 2016). In the present study, VIR data are used to investigate the composition of bright and dark regions, focusing mainly on the major absorptions at ~2.7 µm (phyllosilicate), ~3.3 µm (carbonate and/or organic), and ~3.9 µm (carbonate). Figure 9 displays several VIR IR reflectance spectra of bright and dark material sites including average surface spectra (from HAMO orbit) normalized at 2.59 µm. The displayed typical bright material spectrum (panels A and B) is an average of multiple bright material sites, excluding the unusual bright sites, and also excluding Oxo and Haulani sites. The latter (Oxo and Haulani) have been excluded due to the classification by Palomba et al. (2017) as different families of bright spots. Similarly, the dark material spectrum in Fig. 9A is also an average of several dark material sites (i.e., Occator, Dantu, Urvara, Gaue). Figure 9A demonstrates that the global IR average spectrum is similar to the average bright and dark material spectra, albeit with minor differences in band strengths and spectral slope. It implies that majority of the bright material is composed of Ca-Mg-carbonate, Mg-phyllosilicate with a cerean darkening component (Palomba et al. 2017), while the dark material is found to be similar in composition to the average cerean surface (i.e., Ca-Mg carbonate, Mg-bearing phyllosilicate) but with larger amount of darkening component (e.g., Palomba et al. 2017; De Sanctis et al. 2015). Localized components of water-ice, ground-ice, pitted terrain and activities with potential sublimation and the haze phenomenon are reported in the literature and might affect bright material but are not discussed here (e.g., Combe et al. 2016, 2017; Nathues et al. 2015a, 2017b; Platz et al. 2016; Thangjam et al. 2016b; Sizemore et al. 2017; Schmidt et al. 2017).

Figure 9B shows spectra of Occator-type and Ernutet-type bright material compared to the typical-bright material and the average surface. The ~2.7 µm absorption band minimum of Occator-type is at slightly longer wavelength compared to other spectra, suggesting the presence of an Al-rich phyllosilicate component that is different from the usual cerean Mg-rich phyllosilicate (De Sanctis et al., 2016; Ammannito et al. 2016). The absorption band minimum at ~3.9 µm is also located at longer wavelength than for typical cerean Ca/Mg-rich carbonates, implying Na-rich carbonates (De Sanctis et al. 2016). The VIR spectrum of the Ernutet-type shows rather prominent ~3.4 µm absorptions due to aliphatic organics (De Sanctis et al. 2017).

[Fig 9]

### 4.3 Combined information from FC and VIR

An in-depth analysis of the nature of the bright and dark material on Ceres benefits from the combination of results obtained from VIR and FC data.

[Fig 10]

Figure 10 displays the reflectance at 0.55 µm versus the overall spectral slope represented by the reflectance ratio 0.96/0.44 µm for the bright and dark material. Separate symbols for bright (unfilled circles) and dark (filled circles) material are used. The location of the global average surface is indicated by a '+' symbol. Information on unit compositions, obtained from VIR data, is also included (cf. Palomba et al., 2017; De Sanctis et al., 2015, 2016, 2017; Nathues et al., 2017a). This plot summarizes the overall information from FC and VIR data. The reflectance at 0.55 µm along the x-axis is marked for dark, average surface, and bright material in different shades of gray. The overall slope (0.96/0.44 µm) along the y-axis for the selected sites show a wide range of values. In this parameter space, the bright material at Oxo is seen along a trendline towards the other bright sites (shown by the arrow), and therefore, this cluster represents typical bright material. In section 3.1, we showed that most bright material shows common spectral characteristics in FC wavelengths despite variations in the spectral slope (Figs. 5, 7; peak reflectance at 0.55 µm and blue slope longward). The typical bright material in Fig. 10 shows a trend from very fresh bright material at Oxo to less bright material, e.g., at Kerwan, Ezinu, and Urvara. One may argue that the very bright and fresh material at Oxo and elsewhere is carbonate-rich (Na-bearing, Ca/Mg-bearing carbonate), and thus they are different from other bright material or the average cerean surface. However, spectral data from VIR show that carbonate is ubiquitous on the cerean surface (e.g., De Sanctis et al. 2015). The only difference is the type of carbonate, i.e., Na- or Ca/Mg-, or Mg-carbonate (e.g., De Sanctis et al. 2015, 2016; Palomba et al. 2017). A detailed study about the compositional differences by Palomba et al. (2017) considered Occator, Oxo, and Haulani as separate families of bright spots that are different from their 'typical' bright spot. The 'typical' bright spot (Palomba et al. 2017) and 'typical' bright material defined in the present paper could be different. Meanwhile, the morphology and geologic context of bright material exposed in and around a fresh crater (for example, Oxo, Haulani) shows different types of carbonates that implies a formation during the impact event including post impact modification process or lateral mixing with the dark background material or thermal/aqueous alteration or space weathering in presence of energetic particles. Nevertheless they show a trend of gradual darkening over time.

The data points of Occator-type and Ernutet-type of bright material and their putative trends towards the typical-bright material are illustrated by dashed arrows. Occator-type and Ernutet-type bright materials are distinct from the typical bright material in this parameter space. We note that the data points of spectrally non-reddish bright material at Occator and Ernutet are located within the cluster of the typical bright material data points. The red spectral slope identified in Occator and Ernutet is likely due to distinct compositions and different physical properties of the material. The red slope in Ernutet-type bright material was ascribed to the presence of a considerable amount of aliphatic organics by De Sanctis et al. (2017). However, the cause for the red spectral slope of Occator-type of bright material is still unknown. Sodium carbonate is unlikely to explain this specific spectral property because multiple sites of high sodium carbonate content (e.g., Palomba et al. 2017) do not exhibit an entirely red spectral slope in FC data (e.g., Fig. 5A, E). $NH_4Cl$, NaCl and Al-phyllosilicate are also reported (De Sanctis et al. 2016), however, they are not likely the reddening component of the FC spectra in Occator (e.g., Berg et al. 2016). Though Ahuna-type of bright material is a unique non-impact process, its data points are within the cluster of typical bright material (shown as a triangular symbol in Fig. 10).

The distribution of the data points of bright material and their trend towards the average surface suggest a gradual change by lowering the reflectance towards the average surface over time (indicated by the dotted arrows). Among the bright material, an obvious correlation of age with brightness and spectral slope is observed, but this is not unique throughout. Note that the modal age of the degraded and moderately degraded carters does not necessarily point out the age of the bright material, because most of the bright material in such craters are re-excavated by smaller impacts of primary or secondary origin. Nevertheless, decrease of reflectance of the bright material over time (or darkening effect) is certain. The darkening of the bright material over time could be due to multiple processes including space weathering/impact gardening and mixing with the background surface, accompanied with chemical alteration and thermal metamorphism or shock-blackening changing mineralogical as well as physical properties of the regolith.

Lowering the abundance of the darkening agent of the dark material over time finally reaching the average surface is evident from geologic context and morphology: (1) presence of only some dark material in a few relatively young craters like Occator, Dantu, Urvara, Consus, Gaue and few other localities; (2) paucity of dark material in the majority of the older cratered-terrains and old basins or large craters (except Urvara and Yalode basin); (3) distribution of the dark material data points in the form of a cluster and the location close to the average cerean surface (Fig. 10). It could be because of impact gardening and mixing and alteration or changes in the physical properties of the regolith.

[Fig 11]

Figure 11 is a schematic sketch showing the different types of bright material along with the change of dark and bright material towards the average surface over time. Four types of bright material are shown: (1) typical-bright material of syn-impact processes of different ages and morphology, (2) Ernutet-type of bright material that could be of endogenic or exogenic origin, (3) Occator-type of bright material that is of endogenic and post-impact origin, and (4) Ahuna-type of bright material that is of endogenic and non-impact origin. The typical bright material, i.e., the syn-impact cratering bright material is the most frequent type that is either simply bright due to its fresh morphologic appearance or due to its carbonate-richness (possibly predominantly Na- or Mg/Ca-bearing carbonates of subsurface brine or ice substrates). The formation mechanism and correlation (if any) among the typical bright material of Na- to Ca-Mg-carbonate needs to be investigated in detail. Lateral mixing of the very bright freshly excavated (Na-carbonate rich) with the darker background material might be responsible for the changes accompanied with changes in mineralogical/physical properties. For the dark material that is likely of endogenic origin, it remains unclear whether this is from a dark surface chondritic/primitive material or an impact excavated localized subsurface material. In addition, their spectral and compositional information, and the processes that gradually weakens or removes the darkening component are yet to be studied.

The variety of bright material discussed in this study suggests that the term 'facula' or "faculae" (in plural) need to be used with caution while referring interchangeably to bright spots on Ceres. The 'facula' on planetary surfaces that means 'bright spot' is poorly defined and seldom used but assigned with some sort of unusual characteristics. For example, Sotra Facula on Saturn's moon Titan is believed to be a best example of a cryovolcano (Lopes et al. 2013) including many other faculae. Similarly, Occator bright spots on Ceres are named 'Vinalia Facula' and 'Vinalia Faculae' for the brightest spot and brighter cluster of spots in the crater floor, respectively. However, the Occator bright materials are considered unique having unusual surface composition (an assemblage of ammoniated phyllosilicates/chlorides/carbonate, Na-carbonates, and Al-rich phyllosilicates, in a geologic context consistent with a cryovolcanic origin; De Sanctis et al. 2015, 2016; Palomba et al. 2017; Nathues et al. 2017a, c). Therefore, it is not clear whether 'facula' on Ceres is defined in accordance with either Occator-type of bright spots or every bright spot on Ceres regardless of their composition and formation. Nevertheless, the cerean bright spots might have undergone different evolution processes, and all bright spots are not necessarily be the representative of brine/icy-substrates. Even if they originate from dehydrated brines, they have reached the surface by different processes (excavation due to impact, re-excavation due to impact, extrusion from depth). Therefore, it is important to define faculae and bright materials based on their reflectance, and/or composition, and/or their geologic formation.

## 5.0 Summary

Bright material is ubiquitous on the cerean surface in different geological and morphologic contexts, while dark material is relatively rare. Quantifying the variations and distributions of spectral reflectance properties on dwarf planet Ceres plays a key role in understanding processes that have led to the present surface composition. Our study highlights the importance of defining bright and dark material with respect to the average cerean surface rather than on a relative scale. It is also important to define facula and distinguish from bright material for clarity.

Four types of bright material were defined based on their spectral properties, composition and geologic formation: (1) Typical-bright material that is formed during syn-impact processes and exposed at different ages appearing in different morphologies, (2) Ernutet-type of bright spectrally red material of endogenic or exogenic origin, (3) Occator-type of bright material of endogenic origin, and (4) Ahuna-type of bright material of endogenic origin. Typical bright material is formed during impacts as ejecta or bright cratered material found on crater rims/walls/floors. Though bright ejecta material constitutes the majority of the exposed cerean bright material, it is not necessarily always linked to subsurface briny-/icy-substrates but also could be simply fresh excavates of recent impact that exhibit higher albedo. The origin and formation mechanism of unusual types of bright material are discussed elsewhere: Occator type of bright material formed by ascending subsurface brine material (e.g., Nathues et al. 2017a, c), Ernutet type of bright material being delivered by exogenic projectiles or by endogenic processes (Pieters et al. 2017), Ahuna type of bright material formed by extrusion and inflation of cryomagma (Ruesch et al. 2016; Platz et al. 2017b).

Bright material is found to be composed of Na/Ca-Mg-carbonates, Mg-bearing phyllosilicates with a yet unknown darkening component, while the cerean dark material is found to be similar in composition with the average surface (i.e., Mg-Ca carbonates, Mg-bearing phyllosilicates) but with higher amounts of the darkening component. Whether the cerean dark material is representative of early dark surficial material (primitive chondritic material) or an evolved/processed material from the subsurface that was excavated by impacts remains an open question. The comparative study of bright and dark material with laboratory samples and analogues revealed that physical properties such as grain size, heating/temperature, porosity (compaction) of the regolith need to be considered and quantified while interpreting the spectral properties.

By combining FC and VIR data, we can gain some insights into, and constraints on the nature of the dark and bright material. The normalized VIR spectra show diagnostic spectral differences in the 1-4 µm range. By contrast, the FC data show differences in terms of both reflectance and spectral slope. By plotting various spectral parameters, we found that Ceres dark material may be spectrally homogenous, and its "evolution" to average surface requires an increase in reflectance but little or no change in spectral shape or slope. By contrast, the bright regions show behavior that is more complex including inhomogeneous distribution of endogenous or exogenic origin. Our combined analysis of FC and VIR data of bright and dark material lead us conclude that bright and dark material tends to change gradually towards the average surface over time.

**Acknowledgment**

We thank the Dawn operations team for the development, cruise, orbital insertion, and operations of the Dawn spacecraft at Ceres. Also we would like to thank the Framing Camera operations team, especially P. G. Gutierrez-Marques, J. Ripken, I. Hall and I. Büttner. The Framing Camera project is financially supported by the Max Planck Society and the German Space Agency, DLR. The University of Winnipeg's Planetary Spectrophotometer Facility was established with funding from the Canada Foundation for Innovation, the Manitoba Research Innovations Fund, the Natural Sciences and Engineering Research Council of Canada (NSERC), the Canadian Space Agency (CSA), and the University of Winnipeg (UW), whose support is gratefully acknowledged. We are very thankful to the reviewers N. Stein and Hap McSween (also AE) for their very constructive and helpful comments.

2016. Detection of local H2O exposed at the surface of Ceres. Science 353 (6303), id.aaf3010. Doi: 10.1126/science.aaf3010.

Combe J.-Ph., Raponi A., Tosi F., De Sanctis M. C., Carrozzo F. G., Zambon F., Ammannito E., Hughson K. H. G., Nathues A., Hoffmann M., Platz T., Thangjam G., Schorghofer N., Schröder S. E., Byrne S., Landis M. E., Ruesch O., McCord T. B., Johnson K. E., Singh S. M., Raymond C. A., and Russell C. T., 2017. Exposed H2O-rich areas detected on Ceres with the Dawn Visible and InfraRed mapping spectrometer. Submitted to Icarus. Under review.

Crown, D.A., H. Sizemore, R. Aileen Yingst, S. Mest, T. Platz, D. Berman, N. Schmedemann, D. Buczkowski, D.A. Williams, T. Roatsch, F. Preusker, C. Raymond, C. Russell. Geologic mapping of the Urvara and Yalode Quadrangles, Ceres. Icarus (2017) under review.

De Sanctis M. C., Ammannito E., McSween H. Y., Raponi A., Marchi S., Capaccioni F., Capria M. T., Carrozzo F. G., Ciarniello M., Fonte S., Formisano M., Frigeri A., Giardino M., Longobardo A., Magni G., McFadden L.A., Palomba E., Pieters C.M., Tosi F., Zambon F., Raymond C. A., and Russell C. T. 2017. Localized aliphatic organic material on the surface of Ceres. Science 355 (6326), 719-722. Doi: 10.1126/science.aaj2305.

De Sanctis M. C., Ammannito E., Raponi A., Marchi S., McCord T.B., McSween H. Y., Capaccioni F., Capria M. T., Carrozzo F. G., Ciarniello M., Longobardo A., Tosi F., Fonte S., Formisano M., Frigeri A., Giardino M., Magni G., Palomba E., Turrini D., Zambon F., Combe J.-Ph., Feldman W., Jaumann R., McFadden L.A., Pieters C. M., Prettyman T., Toplis M., Raymond C. A., and Russell C. T. 2015. Ammoniated phyllosilicates with a likely outer Solar System origin on (1) Ceres. Nature 528 (7581), 241-244. DOI: 10.1038/nature16172.

De Sanctis M. C., Coradini A., Ammannito E., Filacchione G., Capria M. T., Fonte S., Magni G., Barbis A., Bini A., Dami M., Ficai-Veltroni I., Preti G., and the VIR Team 2011. The VIR Spectrometer. Space Sci Rev. 163 (1–4), 329-369. Doi: 10.1007/s11214-010-9668-5.

De Sanctis M. C., Raponi A., Ammannito E., Ciarniello M., Toplis M. J., McSween H. Y., Castillo-Rogez J. C., Ehlmann B. L., Carrozzo F. G., Marchi S., Tosi F., Zambon F., Capaccioni F., Capria M. T., Fonte S., Formisano M., Frigeri A., Giardino M., Longobardo A., Magni G., Palomba E., McFadden L. A., Pieters C. M., Jaumann R., Schenk P., Mugnuolo R., Raymond C. A., and Russell C. T. 2016. Bright carbonate deposits as evidence of aqueous alteration on (1) Ceres. Nature 536 (7614), 54-57. Doi: 10.1038/nature18290.

Fries, M., Zolensky, M., Steele, A., 2011, Mineral inclusions in Monahans and Zag halites: Evidence of the originating body. 74th Meteoritical Society Meeting.

Hapke, B., 2012. Theory of Reflectance and Emittance Spectroscopy second ed. Cambridge Univ. Press, Cambridge, UK, (ISBN 978-0-521-88349-8).

Hiesinger, H., Marchi, S., Schmedemann, N., Schenk, P., Pasckert, J.H., Neesemann, A., O'Brien, D.P., Kneissl, T., Ermakov, A.I., Fu, R.R., Bland, M.T., Nathues, A., Platz, T., Williams, D.A., Jaumann, R., Castillo-Rogez, J.C., Ruesch, O., Schmidt, B., Park, R.S., Preusker, F., Buczkowski, D.L., Russell, C.T., Raymond, C.A., 2016. Cratering on Ceres: Implications for its crust and evolution. Science, 353, aaf4759.

Hiroi, T., and M. E. Zolensky (1999), UV-Vis-NIR absorption features of heated phyllosilicates as remote-sensing clues of thermal histories of primitive asteroids, Antarct. Meteorite Res., 12, 108.

**Figure captions:**

Fig. 1: Definition of bright and dark material. From the histogram of reflectance values at 0.55µm using HAMO data (A), a distribution of standard deviations (B, see text) has been calculated. It results in the statistically defined distinction of bright (reflectance >0.035) and dark (reflectance <0.028) material shown in (C). The Data in in Mollweide projection centered at 180°E longitude showing locations of the features that are frequently mentioned in this study. Data in higher latitudes need caution while analyzing and interpreting because of higher uncertainty and noise levels, and are excluded for the statistical analysis. Bright, dark and average surface reflectance are marked in the histogram (A). The statistically defined standard deviation value is marked with a dashed line.

Fig. 2: Global mosaics of Ceres in Mollweide projection centered at 180°E longitude. (A) HAMO/XMO2 color filter mosaic at 0.55 µm. (B) HAMO/XMO2 RGB mosaic combining the color information of three filters (R = 0.96 µm, G = 0.65 µm, B = 0.44 µm). (C) RGB showing spectral slopes expressed by color ratios R = 0.96/0.44 µm, G = 0.65 µm, B = 0.44/0.96 µm. (D) Sites of bright material (yellow dots for craters with diameter < 5 km, green symbols for craters > 5 km) and sites of dark material (red stars) projected on (C).

Fig. 3: Classification of bright material on Ceres based on their geologic context (for explanation see text).

Fig. 4: Bright material on Ceres at selected surface features. The RGB composite images (R= 0.96 µm, G= 0.65 µm, B= 0.44 µm) are derived from HAMO orbit data. Oxo (A), Juling and Kupalo (B), Haulani (C) are fresh appearing, young craters with prominent bright material found on the crater floor/wall/rim/ejecta. Dantu (D), Azacca (E), Ikapati (F) are moderately degraded craters where bright material is found sparsely on crater floor/wall/ejecta when compared to the fresh craters. Ezinu (G), Messor (I), Urvara (H) are degraded craters where the bright material got re-excavated by recent small impacts or by distal ejecta from nearby larger craters. Occator crater (J), being a relatively young crater shows syn-impact bright material as well as post-impact bright material. Ernutet (K) is an eroded crater exhibiting post-impact bright material associated with organic-rich material of yet unknown origin. Ahuna Mons (L) is an exceptional example of non-impact origin, i.e. bright material is associated with a volcanic dome, a result of an endogenic process. All RGB mosaics are stretched equally.

Fig. 5 Reflectance spectra (A-D) and normalized color spectra at 0.65 µm for selected sites of bright material and compared with the average surface. Fresh craters (A, E), moderately-degraded craters (B, F), degraded craters (C, G), unique sites (D, H). The arrows denote the wavelength of peak reflectance at 0.55 µm for bright material (dashed arrow, except for Occator-type and Ernutet-type) and at 0.65 µm for average surface (dotted arrow). The error bars are on the order of the symbol size.

Fig. 6: Display details of the FC HAMO color mosaic at 0.55 µm of prominent dark material sites, Occator dark ejecta in the northeast (A), Nawish/Heneb region (B), Dantu dark ejecta in the southwest (C), parts of Urvara dark floor/ejecta (D). Images are stretched at the same scale. Reflectance spectra (E) and normalized spectra at 0.65 µm (F) for selected sites of dark material compared with the global average. Dotted arrows denote the wavelength of peak reflectance at 0.65 µm. Error bars are on the order of the symbol size.

Fig. 7: Average bright and dark material color spectra (A) and normalized reflectance spectra at 0.96 µm (B) on Ceres compared with the global average. The arrows denote peak reflectance at 0.55 µm for bright material (dashed arrow) and at 0.65 µm for dark material and average surface (dotted arrow). Typical bright material and unusual bright material reflectance spectra (C) and normalized spectra at 0.96 µm (D) compared with the average surface. Because of the high reflectance of Occator bright-red material, a second y-axis is plotted (C). The error bars are in the order of the symbol size.

Fig. 8: Overall spectral slope 0.96/0.44µm versus reflectance at 0.55 µm (A-D), and spectral slope 0.96/0.65 µm versus spectral slope 0.44/0.65 µm (E-H), for cerean bright (typical-bright and unusual bright) and dark material compared to various laboratory samples/analogues in different physical states and properties, (A, E) grain size dependences, (B, F) fineness/coarseness dependence, (C, G) heating/temperature dependence, and (D, H) compactness or porosity dependence. Details on grain sizes ranges and heating temperatures are given in the text.

Fig. 9: VIR IR spectra normalized at 2.59 µm of selected bright and dark material sites compared with the global average Ceres spectrum from HAMO orbit. Spectral gaps are introduced by order-sorting filter calibration errors and data artifacts. The absorption features at ~3.4 and ~3.9 µm are indicative of carbonates, ~3.4 µm for organics, and ~2.7 µm for phyllosilicates.

Fig. 10: Overall spectral slope (0.96/0.44µm) versus reflectance at 0.55 µm for bright (open circles) and dark (filled circles) material along with the average surface (plus symbol) in logarithmic scale of base 10. The data point with triangular symbol is for Ahuna Mons. Reflectance of dark material, average surface, and bright material are indicated in different shades of gray. Trendline of typical-bright material, and likely a gradual change of Occator-type of bright material and Ernutet-type of bright material towards the typical bright material are shown with the arrows. Dotted arrows suggest a gradual change of bright material towards the global average surface over time. Cluster of dark material close towards the average surface and their morphologic distribution suggest a gradual removal or weakening of the darkening component. The given endmember information (in italics) is in accordance with the VIR observations (e.g., De Sanctis et al. 2015, 2016, 2017; Palomba et al. 2017).

Fig. 11 Bright and dark material over time that tends to change gradually towards the average surface. Four types of bright material are illustrated: Typical-bright material of syn-impact processes of different ages and morphology, Ernutet-type of bright spectrally red material of endogenic or exogenic origin, Occator-type of bright spectrally red material of endogenic origin, and Ahuna Mons type of bright material of endogenic origin.

SOM/S1: Bright material on Ceres at selected surface features. The RGB composite images (R= 0.96 µm, G= 0.65 µm, B= 0.44 µm) are derived from HAMO orbit data. Oxo (A), Juling and Kupalo (B), Haulani (C) are fresh appearing, young craters with prominent bright material found on the crater floor/wall/rim/ejecta. Dantu (D), Azacca (E), Ikapati (F) are moderately degraded craters where bright material is found sparsely on crater floor/wall/ejecta when compared to the fresh craters. Ezinu (G), Messor (I), Urvara (H) are degraded craters where the bright material got re-excavated by recent small impacts or by distal ejecta from nearby larger craters. Occator crater (J), being a relatively young crater shows syn-impact bright material as well as post-impact bright material. Ernutet (K) is an eroded crater exhibiting post-impact bright material associated with organic-rich material of yet unknown origin. Ahuna Mons (L) is an exceptional example of non-impact origin, i.e. bright material is associated with a volcanic dome, a result of an endogenic process. All RGB mosaics are stretched differently.

SOM/Catalogue of bright and dark material

**Fig. 1:**

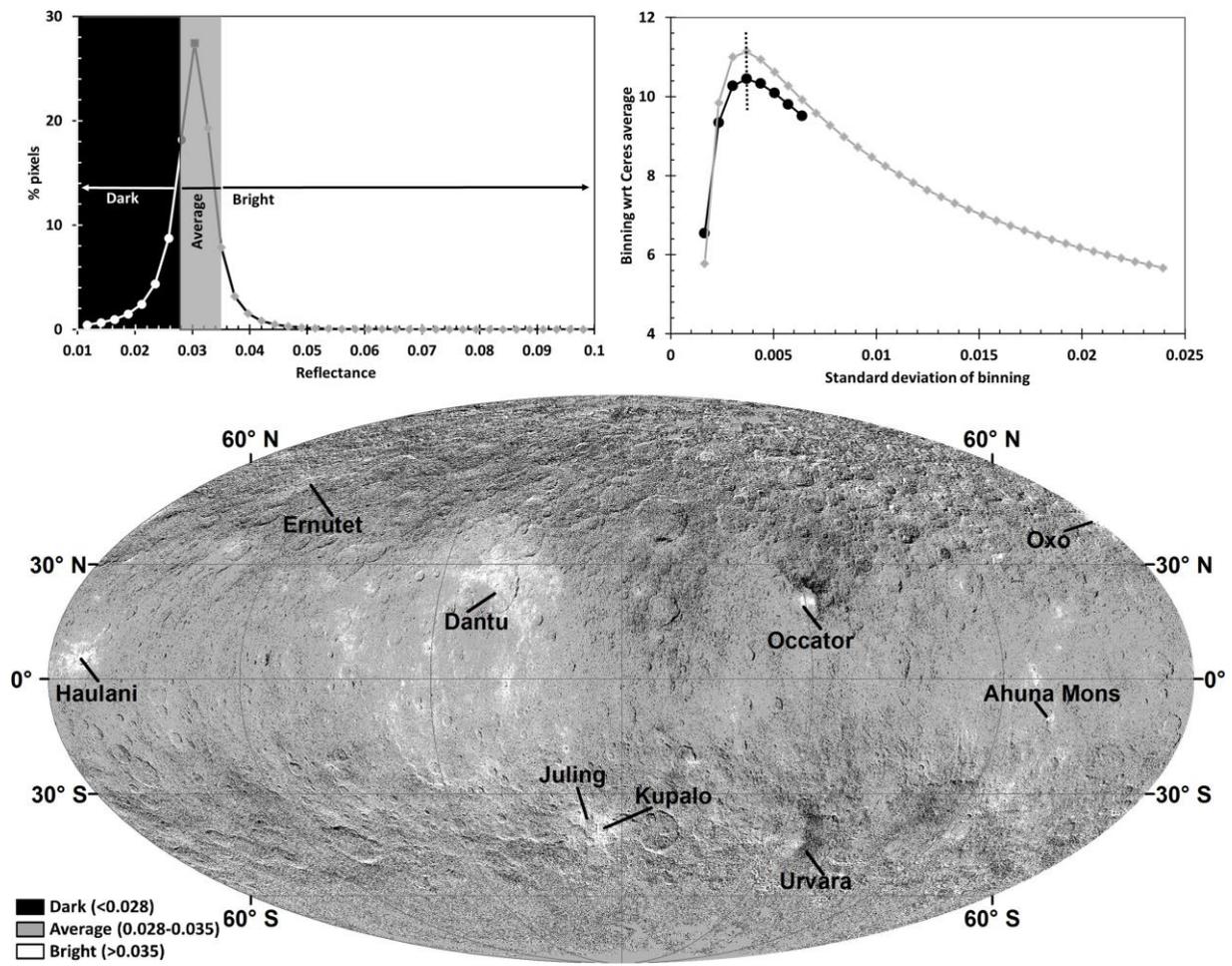

Fig. 2:

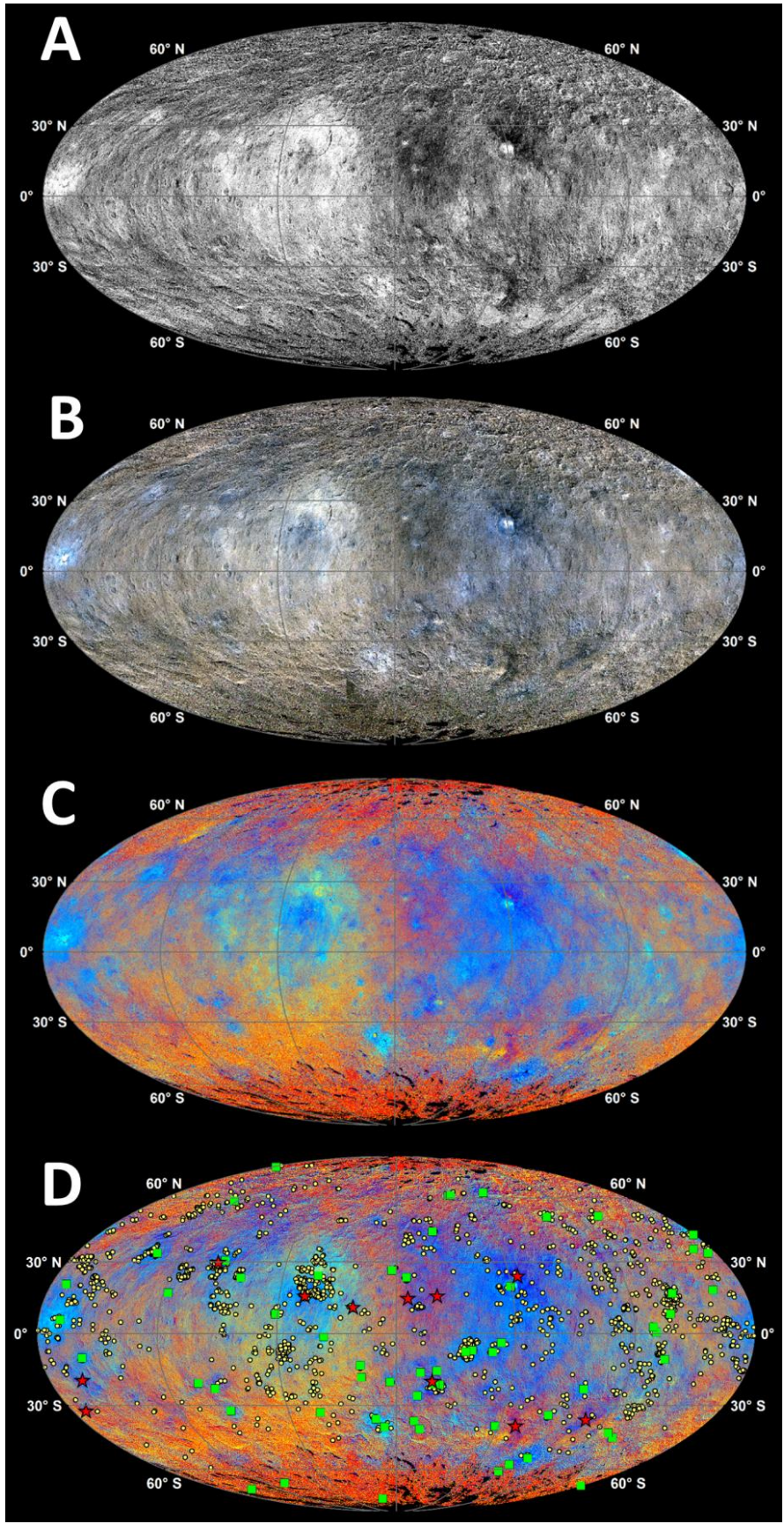

Fig. 3:

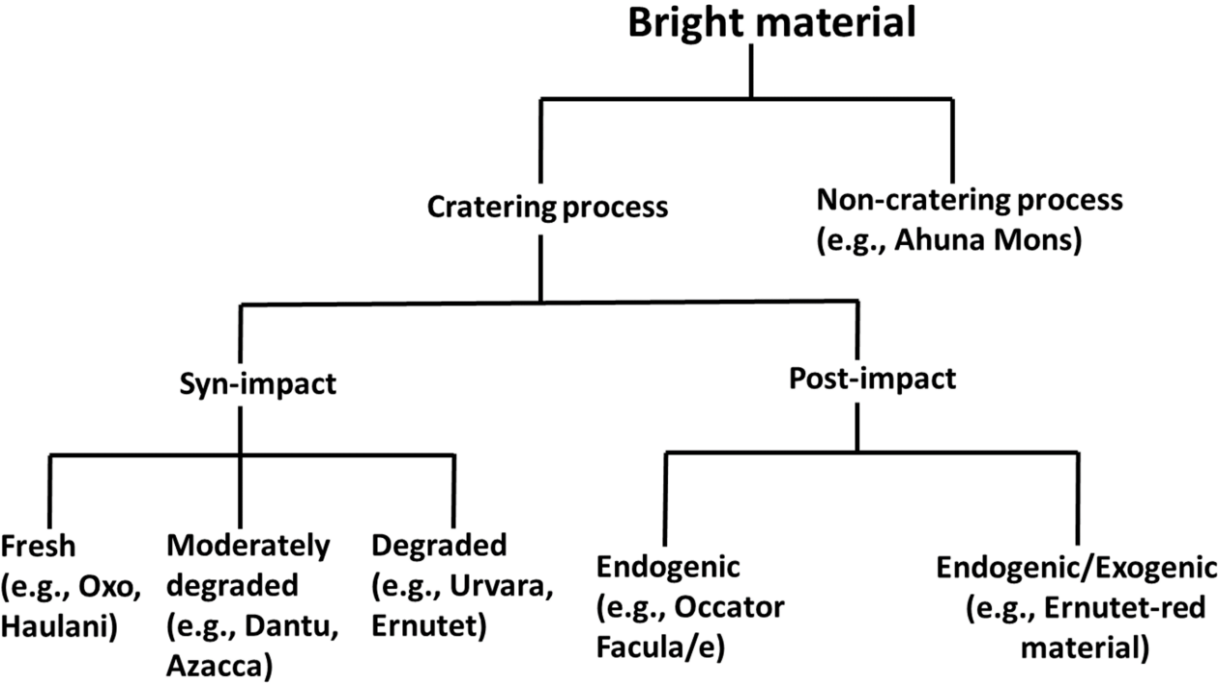

Fig. 4:

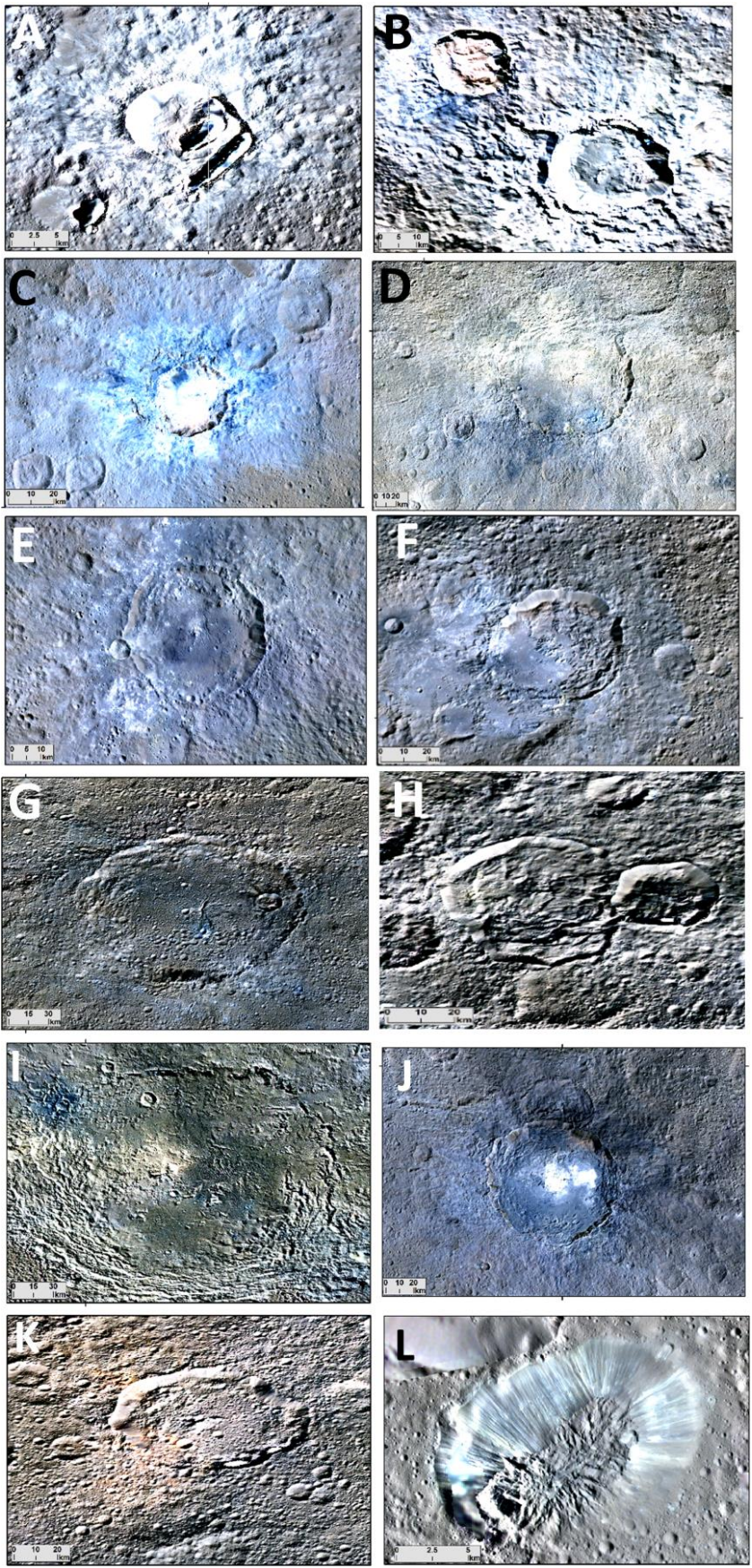

Fig. 5:

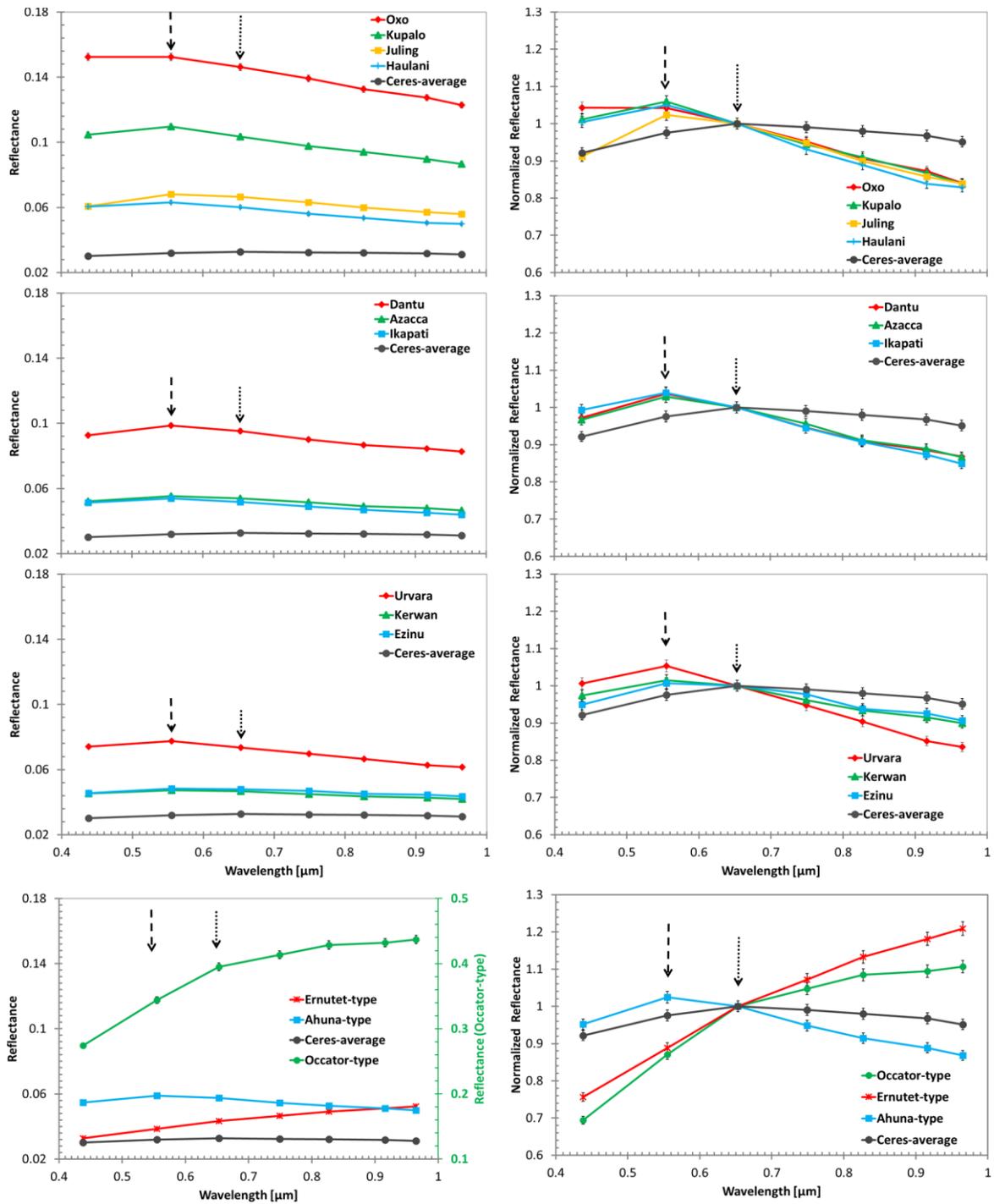

Fig. 6:

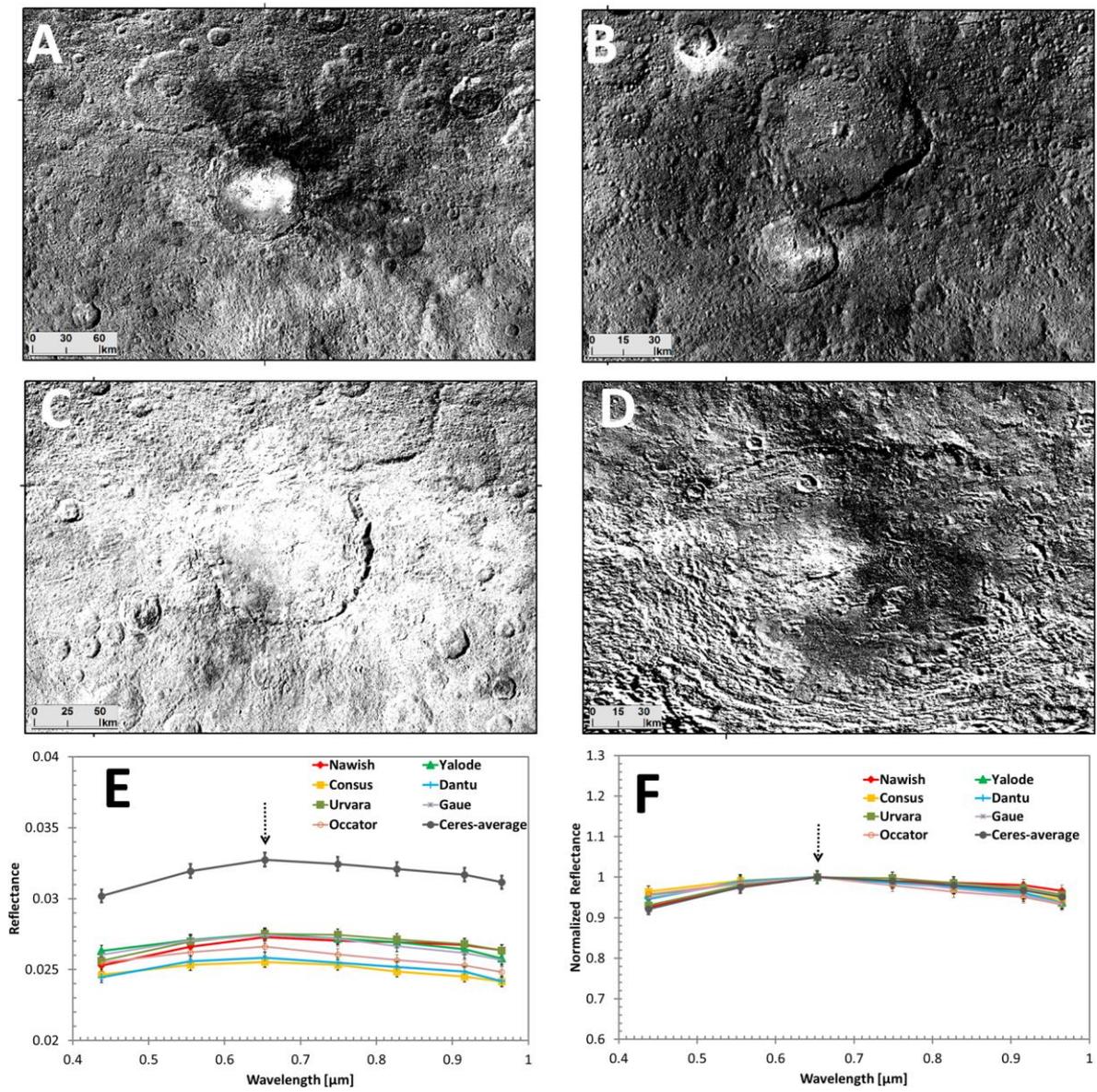

Fig. 7:

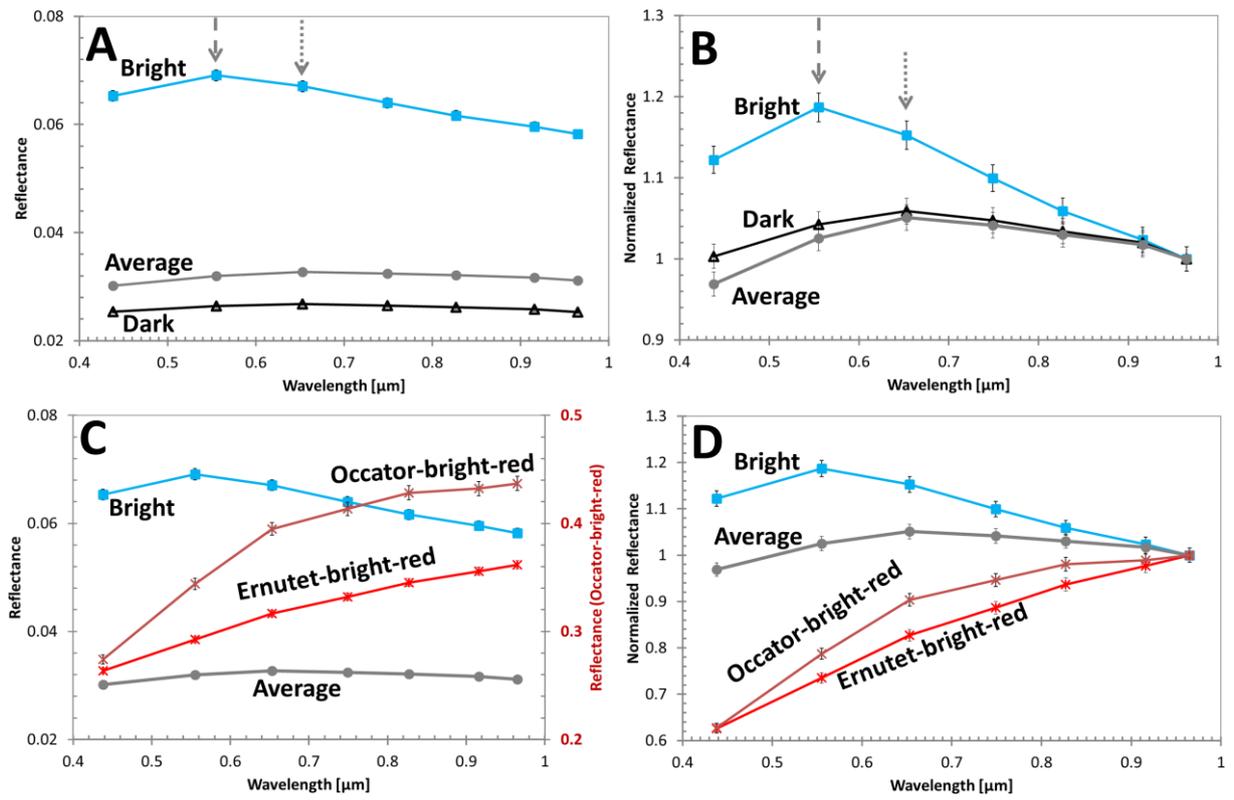

Fig. 8:

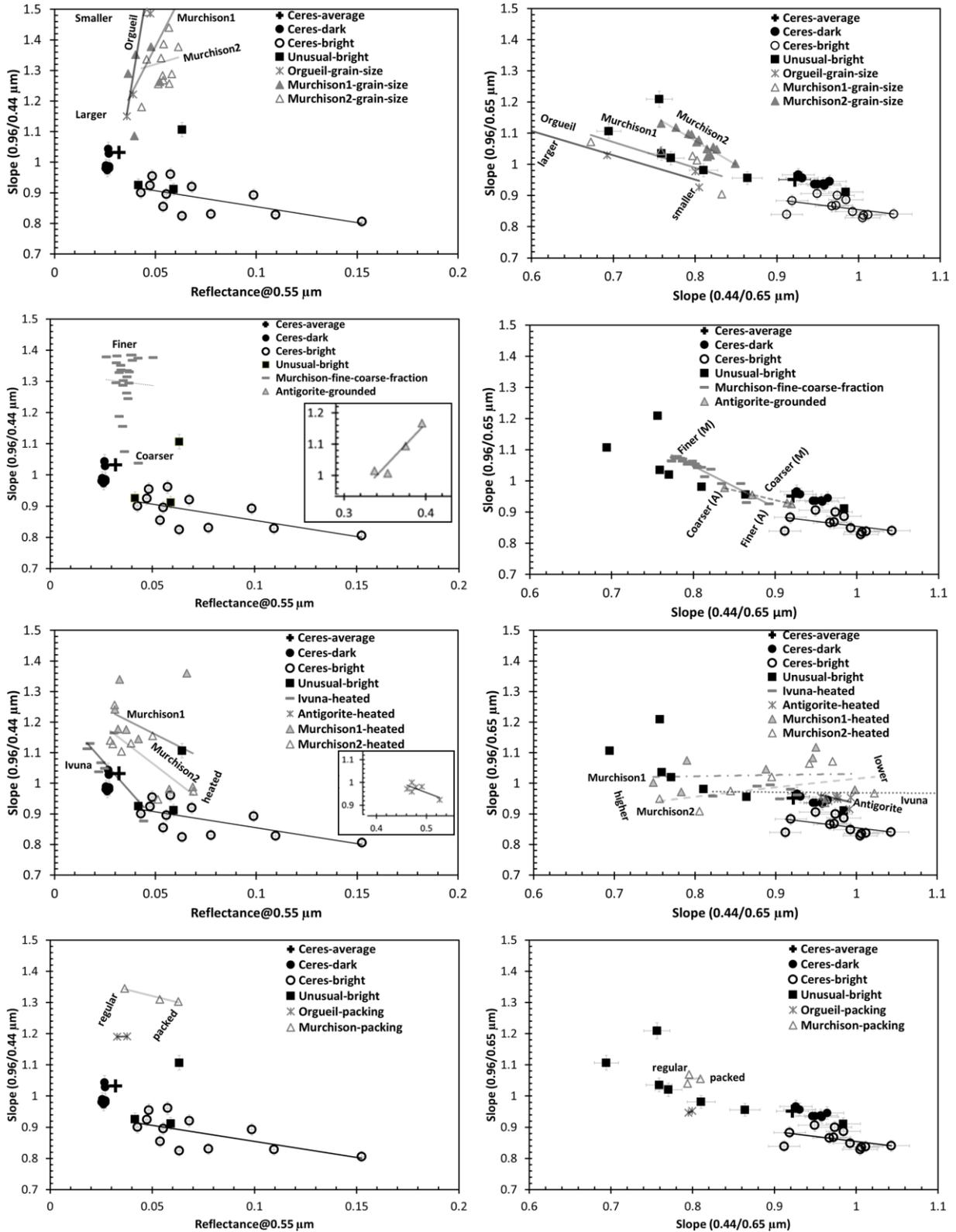

Fig. 9:

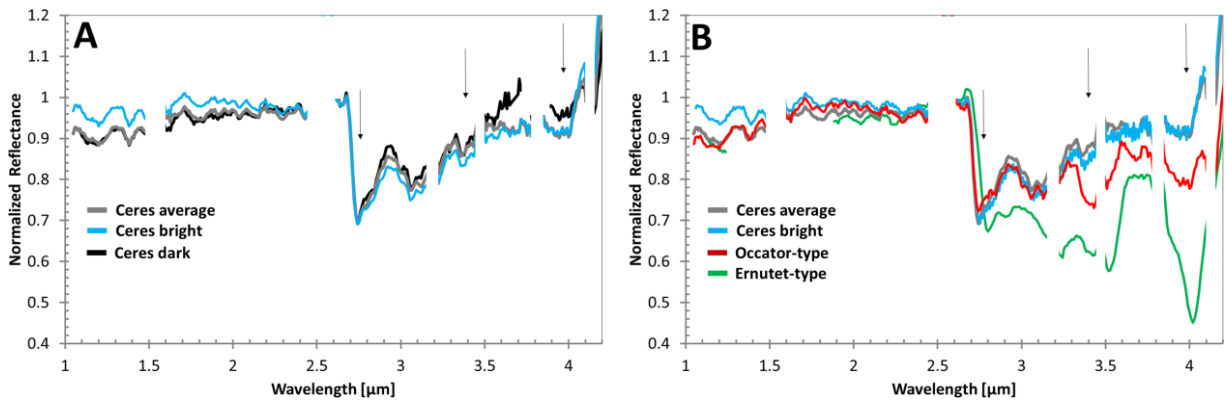

Fig. 10:

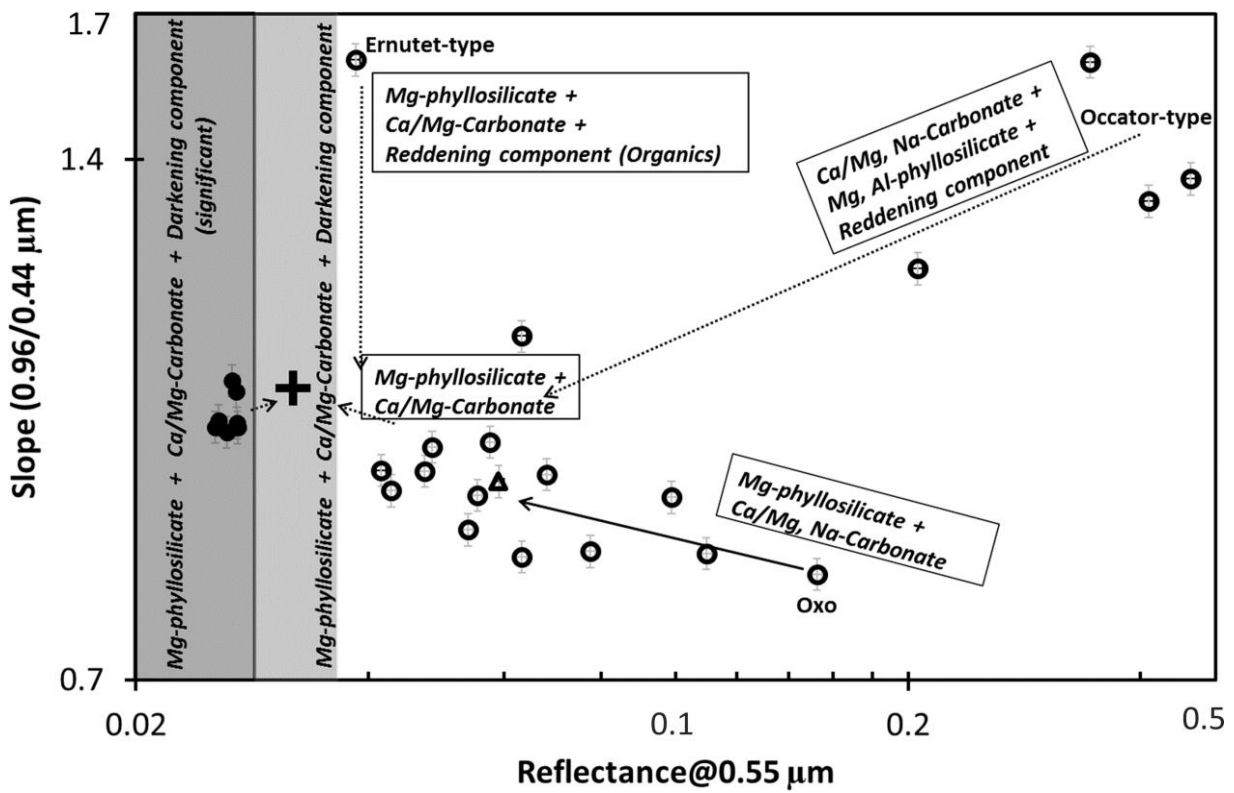

Fig. 11:

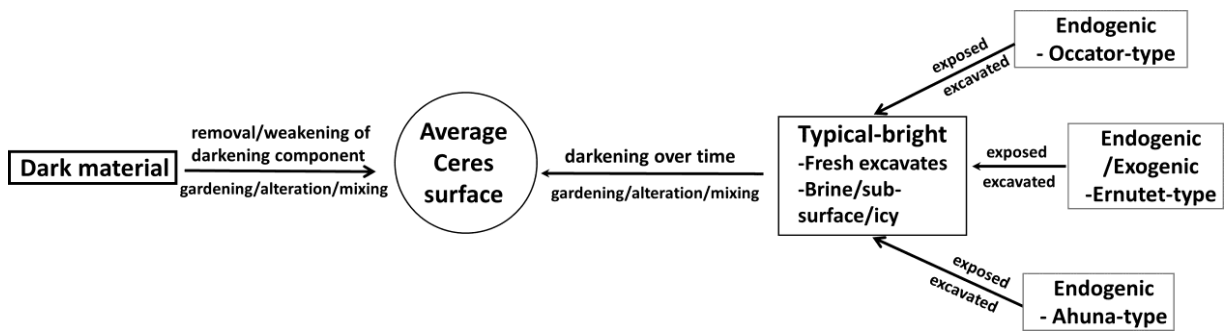

Fig. S1:

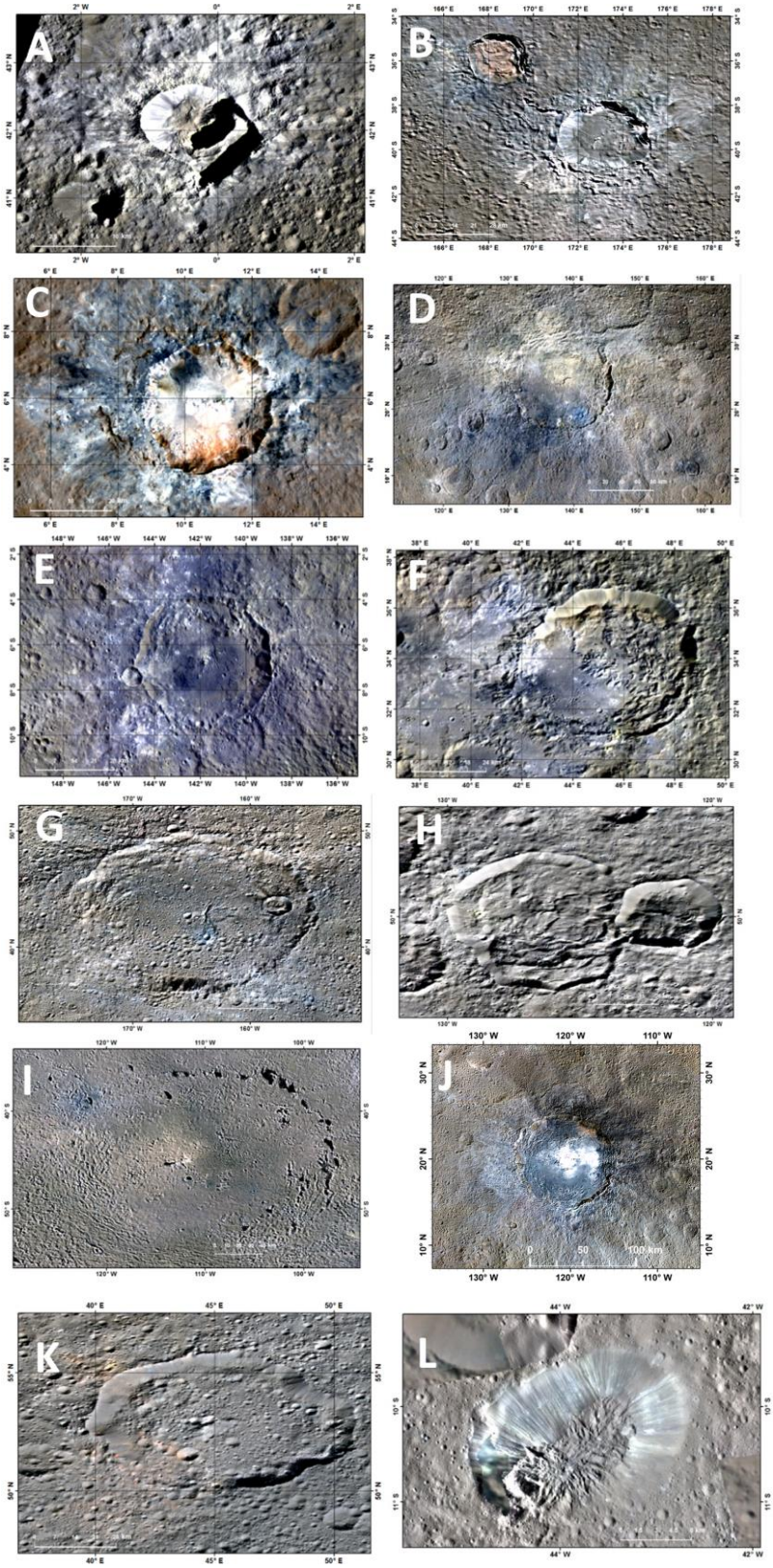

**Catalogue of Bright material:**

| Sl. No. | Longitude | Latitude | Feature-name |
|---|---|---|---|
| 1 | -10.14 | 35.68 | Unnamed crater, West of Duginavi crater |
| 2 | -0.37 | 42.09 | Oxo crater in Duginavi crater |
| 3 | -37.70 | 16.63 | Begbalel crater |
| 4 | -41.31 | 8.23 | Unnamed crater, NE of Liberalia Mons |
| 5 | -49.41 | 0.62 | Unnamed crater, south of Liberalia Mons |
| 6 | -50.60 | 2.78 | Unnamed crater, south of Liberalia Mons |
| 7 | -157.17 | -21.07 | Unnamed crater in Consus crater |
| 8 | -141.53 | -6.76 | Azacca crater |
| 9 | -144.79 | -7.42 | Unnamed crater, west of Azacca crater |
| 10 | 173.20 | -39.44 | Kupalo crater |
| 11 | 168.51 | -35.89 | Juling crater |
| 12 | 137.63 | -32.98 | Unnamed crater, SE of Kerwan crater |
| 13 | 76.62 | -20.56 | Unnamed crater, NW of Braciaca crater |
| 14 | 88.14 | -32.31 | Tupo crater |
| 15 | 10.78 | 5.76 | Haulani crater |
| 16 | 7.72 | 20.52 | Unnamed crater, south of Duginavi crater |
| 17 | 45.61 | 33.83 | Ikapati crater |
| 18 | 5.96 | 79.06 | Unnamed crater, SW of Yamor Mons |
| 19 | -120.33 | 19.64 | Occator crater |
| 20 | -138.55 | 61.32 | Unnamed crater, NW of Messor crater |
| 21 | -136.54 | 61.53 | Unnamed crater, NW of Messor crater |
| 22 | -108.06 | 62.97 | Unnamed crater, North of Datan crater |
| 23 | -79.64 | 50.49 | Takel crater |
| 24 | 119.01 | 8.11 | Rao crater |
| 25 | 77.42 | -67.49 | Unnamed crater, SW of Hamori crater |
| 26 | 162.20 | -77.70 | Unnamed crater, SE of Chaminuka crater |
| 27 | -0.56 | -69.34 | Unnamed crater, near Mondamin crater (SW) |
| 28 | -43.78 | -10.51 | Ahuna Mons |
| 29 | 138.68 | 24.06 | Dantu crater |
| 30 | 97.83 | 23.25 | Arisaeus crater |
| 31 | 86.01 | 30.51 | Gaue crater |
| 32 | 20.99 | -9.95 | Unnamed crater, NE of Anura crater |
| 33 | -121.96 | -39.13 | Tawals crater |
| 34 | -46.14 | -44.41 | Unnamed crater, east of Yalode crater |
| 35 | -15.94 | 18.07 | Unnamed crater, east of Begbalel crater |
| 36 | -4.40 | 33.82 | Unnamed crater, in Duginavi crater (SW) |
| 37 | -174.46 | 23.41 | Unnamed crater, NW of Nawish crater |
| 38 | 177.69 | 26.35 | Unnamed crater, NW of Nawish crater |
| 39 | 143.62 | -1.19 | Cacaughat crater |
| 40 | 84.41 | -22.74 | Braciaca crater |
| 41 | -87.71 | -54.19 | Unnamed crater, southeast of Urvara crater |

| 42 | -96.19 | -57.44 | Unnamed crater, southeast of Urvara crater |
|---|---|---|---|
| 43 | -99.21 | -60.92 | Unnamed crater, southeast of Urvara crater |
| 44 | 30.79 | -71.40 | Unnamed crater, west of Zadeni crater |
| 45 | 57.94 | 58.54 | Unnamed crater, west of Omonga crater |
| 46 | -80.95 | -22.91 | Unnamed crater, northwest of Yalode crater |
| 47 | -93.91 | -34.32 | Unnamed crater, northwest of Pongal Catena |
| 48 | 176.78 | -19.92 | Unnamed crater, south of Fluusa crater |
| 49 | 161.81 | -13.07 | Unnamed crater, southwest of Fluusa crater |
| 50 | 162.08 | -18.02 | Unnamed crater, southwest of Fluusa crater |
| 51 | 62.36 | 16.98 | Unnamed crater, southwest of Achita crater |
| 52 | -157.58 | 43.57 | Unnamed crater, in Ezinu crater (west) |
| 53 | -44.60 | 50.95 | Unnamed crater, northwest of Hosil Tholus |
| 54 | -51.79 | -42.07 | Unnamed crater, west of Besua crater |
| 55 | -167.31 | -15.89 | Unnamed crater, northwest of Consus crater |
| 56 | -159.09 | -15.30 | Unnamed crater, north of Consus crater |
| 57 | -168.59 | -25.81 | Unnamed crater, southwest of Consus crater |
| 58 | -159.67 | -20.83 | Consus crater |
| 59 | -169.72 | -36.82 | Unnamed crater in Meanderi crater (northwest) |
| 60 | -166.16 | -40.46 | Meanderi crater |
| 61 | -127.13 | -3.71 | Unnamed crater, northeast of Lociyo crater |

**Catalogue of dark material:**

| Sl. No. | Longitude | Latitude | Feature name |
|---|---|---|---|
| 1 | -110.15 | -38.89 | Urvara crater |
| 2 | -115.53 | 24.13 | Occator crater |
| 3 | -174.03 | 14.66 | Nawish/Heneb crater |
| 4 | -158.98 | 15.74 | Nawish/Heneb crater |
| 5 | -71.44 | -36.10 | Yaolde crater |
| 6 | 133.06 | 15.74 | Dantu crater |
| 7 | 82.51 | 29.94 | Gaue crater |
| 8 | -161.20 | -19.55 | Consus crater |
| 9 | 158.02 | 11.07 | Unnamed crater, SE of Dantu |
| 10 | 7.31 | -32.25 | Unnamed crater, S of Wangala Tholus |
| 11 | 16.89 | -19.08 | Kandos crater |